\documentclass[10pt,aps,prd,nofootinbib,superscriptaddress,twocolumn,preprintnumbers,balancelastpage]{revtex4}

\usepackage{amssymb,amsmath,latexsym,graphics, graphicx,epsfig,multirow,comment,appendix,feyn,slashed,color,afterpage,multirow,makecell} 

\usepackage[colorlinks=true
,urlcolor=blue
,anchorcolor=blue
,citecolor=blue
,filecolor=blue
,linkcolor=blue
,menucolor=blue
,linktocpage=true
,pdfproducer=medialab
,pdfa=true
]{hyperref}

\usepackage{graphicx}
\usepackage{lipsum}
\usepackage{bm}
\usepackage{setspace}

\definecolor{colorTC}{rgb}{.2,.7,.2}

\newcommand{\s}{\hspace{0.8pt}}

\newcommand{\be}{\begin{equation}}
\newcommand{\ee}{\end{equation}}
\newcommand{\bea}{\begin{eqnarray}}
\newcommand{\eea}{\end{eqnarray}}
\newcommand{\beq}{\begin{equation}}
\newcommand{\eeq}{\end{equation}}

\def\Eq#1{eq.~(\ref{#1})}
\def\Fig#1{fig.~\ref{#1}}
\def\Sec#1{sect.~\ref{#1}}

\newcommand{\LUV}{\Lambda_\text{UV}}

\def\Eq#1{Eq.~(\ref{#1})}
\def\Fig#1{Fig.~(\ref{#1})}
\def\Sec#1{Sect.~(\ref{#1})}

\def\U{\text{U} }

\begin{document}
%%%%%%%%%%%%%%%%%%%%%%%%%%%%%%%%%%%%%%%%%%%%%%%%%%%%%%%%%%%%%%%%%%%%%%
%\preprint{ULB-TH/17-??}
\preprint{CERN-TH-2020-013}
%\preprint{DAMTP-2019-X}
%{\raggedleft DAMTP-2019-X \\}

%\title{Logging a natural UV-conspiracy}
%\title{Logging in the UV to Yield a Natural Theory}
%\title{Logging Away the UV Yields a Natural Weak Scale}
%\title{UV Logging to Yield a Natural Theory}
%\title{The Weak Scale Logging out of UV-Sensitivity}}
%\title{Logging Out the Weak Scale}
\title{Supersoft Stops}

\author{Timothy Cohen}
\affiliation{Institute for Fundamental Science, Department of Physics, University of Oregon, Eugene, OR 97403, USA\vspace{2pt}}

\author{Nathaniel Craig}
\affiliation{Department of Physics, University of California, Santa Barbara, CA 93106, USA\vspace{2pt}}

\author{Seth Koren}
\affiliation{Department of Physics, University of California, Santa Barbara, CA 93106, USA\vspace{2pt}}

\author{Matthew McCullough}
\affiliation{CERN, Theoretical Physics Department, Geneva, Switzerland\vspace{2pt}}
\affiliation{DAMTP, University of Cambridge, Wilberforce Road, Cambridge, UK\vspace{2pt}}

\author{Joseph Tooby-Smith}
\affiliation{Cavendish Laboratory, University of Cambridge, Cambridge, UK\vspace{2pt}}

%%%%%%%%%%%%%%%%%%%%%%%%%%%%%%%%%%%%%%%%%%%%%%%%%%%%%%%%%%%%%%%%%%%%%%
\begin{abstract}
\begin{center}
{\bf Abstract}
\vspace{-7pt}
\end{center}
In a supersymmetric (SUSY) theory, the IR-contributions to the Higgs mass are calculable below the mediation scale $\Lambda_{\text{UV}}$ in terms of the IR field content and parameters. 
However, logarithmic sensitivity to physics at $\Lambda_{\text{UV}}$ remains.
In this work we present a first example of a framework, dictated by symmetries, to supersoften these logarithms from the matter sector.
The result is a model with finite, IR-calculable corrections to the Higgs mass.
This requires the introduction of new fields -- the `lumberjacks' -- whose role is to screen the UV-sensitive logs.
These models have considerably reduced fine-tuning, by more than an order of magnitude for high scale supersymmetry.  
This impacts interpretations of the natural parameter space, suggesting it may be premature to declare a naturalness crisis for high-scale SUSY.
\end{abstract}
%%%%%%%%%%%%%%%%%%%%%%%%%%%%%%%%%%%%%%%%%%%%%%%%%%%%%%%%%%%%%%%%%%%%%%

%%%%%%%%%%%%%%%%%%%%%%%%%%%%%%%%%%%%%%%%%%%%%%%%%%%%%%%%%%%%%%%%%%%%%%
\maketitle
%%%%%%%%%%%%%%%%%%%%%%%%%%%%%%%%%%%%%%%%%%%%%%%%%%%%%%%%%%%%%%%%%%%%%%

%\begin{spacing}{1.07}

%%%%%%%%%%%%%%%%%%%%%%%%%%%%%%%%%%%%%%%%%%%%%%%%%%%%%%%%%%%%%%%%%%%%%%
%\section{Introduction}
%\label{sec:intro}
%%%%%%%%%%%%%%%%%%%%%%%%%%%%%%%%%%%%%%%%%%%%%%%%%%%%%%%%%%%%%%%%%%%%%%

The search for models of short distance physics that yield a calculable Higgs mass continues to be one of the most important motivations for explorations beyond the Standard Model.
This reductionist approach to fundamental questions has many precedents.
The scalar mass in the Landau-Ginzburg theory is calculable utilizing the more microscopic BCS theory of superconductivity.  
Hadron masses derive from the underlying QCD quark mass and gauge coupling parameters.  
Yet, despite our best efforts, no experimental hints towards the provenance of the Higgs mass have emerged, leading to the commonly held belief that the Standard Model must be fine-tuned.
Here, via a novel extension of the Standard Model matter sector, we challenge the notion that the absence of evidence for naturalness (thus far) is evidence of its absence.
 
Taking seriously the notion that the Higgs mass parameter $M_H$ originates at some microscopic scale `$\LUV$', along with the observed lack of protective symmetries within the Standard Model, results in the expectation that $M_H$ should be parametrically determined by $\LUV$.  
Scenarios that can accommodate a comfortable separation $M_H \ll \LUV$ rely on invoking additional symmetries, which imply the existence of associated fields that are required to fill out complete representations.  
Since the top quark has the largest coupling to the Higgs boson, one expects that the relationship between $M_H$ and $\LUV$ is dominated by new physics in the top sector.
The best studied examples are supersymmetry (SUSY), which introduces the coloured scalar stop squarks, or spontaneously broken global symmetries, which requires a set of coloured fermionic top partner fields.  
If the mass scale of such fields is $\widetilde{m}$ then the underlying dependence of the Higgs mass on the UV physics at $\LUV$ is softened from a quadratic sensitivity $\delta M_H^2 \propto \LUV^2$, to a logarithmic one $\delta M_H^2 \propto \widetilde{m}^2/(16\s \pi^2) \log (\LUV/\widetilde{m})$.  
This remaining logarithmic sensitivity to $\LUV$ betrays the fact that although $\widetilde{m} \ll \LUV$ can be naturally accommodated, the Higgs mass is only truly calculable if one knows the complete theoretical picture at $\LUV$. 
The implications of this residual UV dependence for the naturalness of the weak scale can be profound, heightening the sensitivity of the Higgs mass to the scale of symmetry breaking.

\begin{figure}[t]
\begin{center}
\includegraphics[width=0.95\columnwidth]{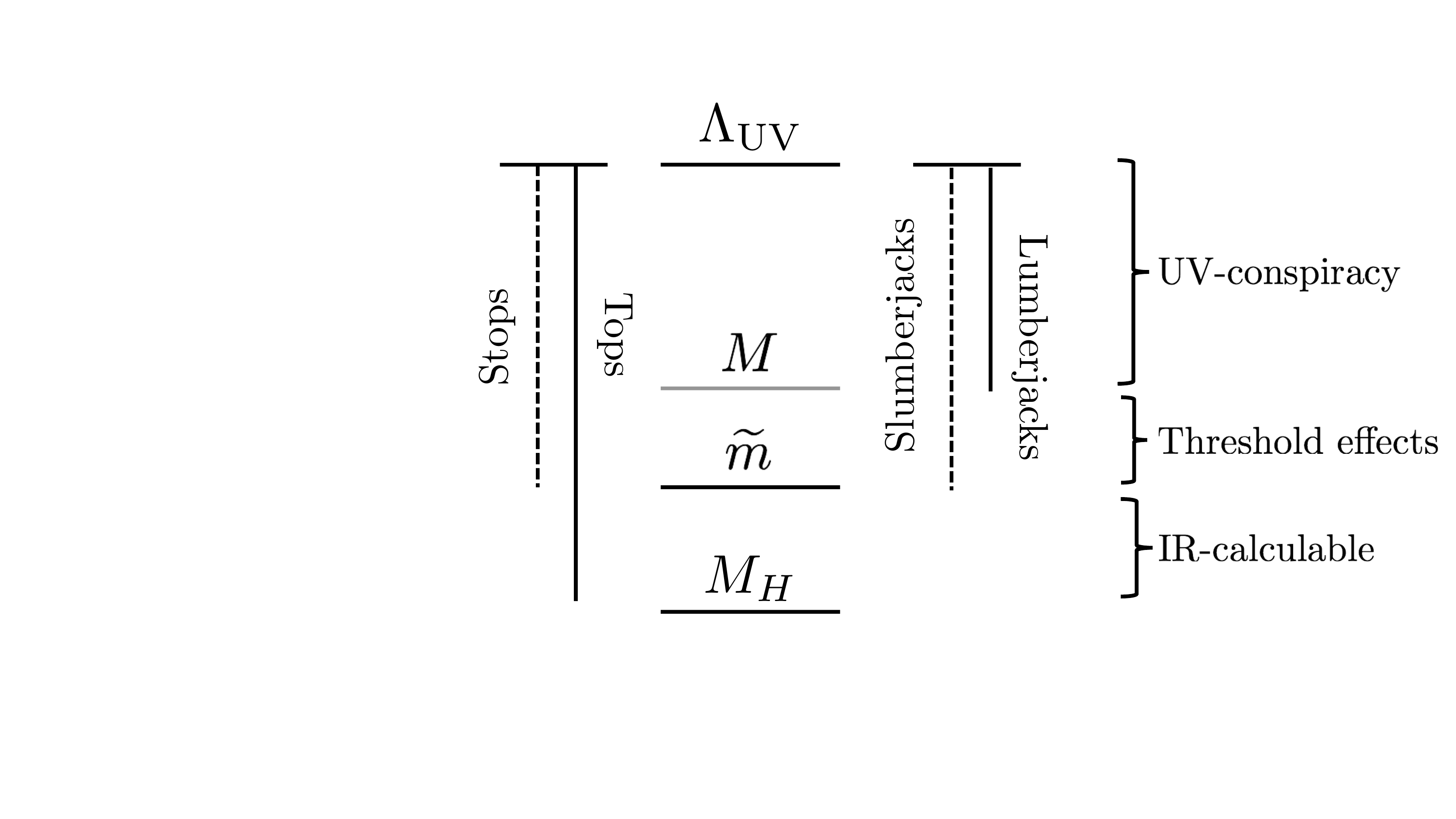}
\end{center}
\caption{A sketch of the scales relevant to the broad picture, with the running contributions depicted on either side of the mass scales.  Above the scale $M$ the lumberjack fields conspire with the MSSM states to screen running Higgs mass corrections from the UV-scale $\LUV$ all the way down to the IR.  Calculable threshold corrections to the Higgs mass are generated below the mass scale $M$.  Scalars (dashed) are separated from their partner fermions (solid) by an equal and opposite-signed soft-mass squared.}
\label{fig:idea}
\end{figure}

In this work, we investigate to what extent the scaling $\delta M_H^2 \propto \widetilde{m}^2/(16\s\pi^2) \log (\LUV/\widetilde{m})$ is inevitable.  
Instead of demanding that the mass scale $\widetilde{m}$ is small to preserve naturalness, we reorient the question by instead asking if it is possible that the logarithm can be rendered insensitive to the UV scale.
Specifically, we will introduce scenarios in which an \emph{additional} symmetry-enforced cancellation occurs over a large range of scales, essentially screening the Higgs mass from $\widetilde{m}$-dependent corrections all the way from the microscopic scale $\LUV$ (where the symmetry breaking is generated) down to the TeV-scale.  
Corrections are still present at all scales, as expected for logarithmic UV-sensitivity.
However, a UV conspiracy enforces an additional cancellation to remove logs, a phenomenon we will call `logging'.
In practise, this amounts to creating an equal-and-opposite `little anti-hierarchy problem', that largely annihilates the little hierarchy problem typically present in SUSY models.  
The framework is depicted in \Fig{fig:idea}.

Importantly, logging away all UV-sensitivity implies that the Higgs mass corrections must be fully calculable in terms of IR parameters.  
Some examples of IR determined contributions to the Higgs mass already exist in the literature.  
In perhaps the most familiar class of models, of which logging is a member, symmetries orchestrate screening among the IR degrees of freedom.
For example, in `supersoft' theories with Dirac Gauginos, the one-loop gauge corrections to scalar masses are finite \cite{Fox:2002bu} -- one way to view the model presented here is that it demonstrates how to supersoften the matter sector. 
In maximally symmetric pNGB Higgs models \cite{Csaki:2017cep}, a remnant symmetry renders the Higgs potential finite.
A second class of models are those in which there are simply no local operators that could contribute to the Higgs mass at $\LUV$, and hence no logarithms are associated with their evolution.
In models of Scherk-Schwarz SUSY breaking, locality ensures that all scalar mass corrections are finite and calculable \cite{Scherk:1979zr,Scherk:1978ta}, albeit with coefficients that are relatively large with respect to the mass of the lightest coloured state.  
In Little Higgs models \cite{ArkaniHamed:2001nc,ArkaniHamed:2002pa}, collective symmetry breaking and the attendant non-locality in theory space results in one-loop finite mass corrections.  
Alternatively, one can mimic IR domination by considering UV-sensitive models with a very low $\LUV$, which by design implies small logarithms, yielding reduced fine-tuning for the viable parameter space.

A common aspect of the EFT description of many of these IR-dominated models is that symmetries and/or locality forbid the possibility of local counterterms which correct the Higgs mass, rendering the Higgs mass IR-calculable.   
Motivated by this observation, our purpose here is to introduce a new class of supersymmetric effective field theories, in which the interplay of global exchange symmetries and symmetry breaking forbid these local counterterms.
The result will be that incalculable logarithms are screened, such that weak-scale fine-tuning from the top sector is significantly reduced.

\setcounter{section}{1}
\section{Logging SUSY}
\label{sec:Scalars}
%%%%%%%%%%%%%%%%%%%%%%%%%%%%%%%%%%%%%%%%%%%%%%%%%%%%%%%%%%%%%%%%%%%%%%
In this section, we will present a low energy SUSY EFT with supersoft stops.
We add a complete copy of the third generation matter, the `lumberjack' fields whose purpose is to remove the logs.\footnote{One could duplicate all of the MSSM matter in the same way, with a minimal impact on the resulting tuning.}  
These lumberjacks are related to the third generation matter by an exchange symmetry.  
As a result, the most general matter-sector renormalizable superpotential is
\begin{align}
\boldsymbol{W_\lambda} = \lambda_t\, \boldsymbol{H_u}\, \boldsymbol{Q}\, \boldsymbol{U^c} + \lambda_t\, \boldsymbol{H_u}\, \boldsymbol{Q'}\, \boldsymbol{U^{\prime c}}~~.
\label{eq:Wlambda}
\end{align}
In order to lift the lumberjack fields, we will pair them up with an additional set of vector-like partners.  
A non-zero vector-like mass $M$ softly breaks the exchange symmetry:
\be
\boldsymbol{W_M} = M \left( \boldsymbol{Q\strut^{\prime}}\, \boldsymbol{\overline{Q}\strut^{\prime}} +\boldsymbol{U\strut^{\prime c}}\, \boldsymbol{\overline{U}\strut^{\prime c}} \right)~~.
\label{eq:WM}
\ee
Note that the exchange symmetry is not broken by the spectrum itself, since $\boldsymbol{\overline{Q}\strut^{\prime}}$ and $\boldsymbol{\overline{U}\strut^{\prime c}}$ do not transform.

The matter squarks and their scalar lumberjack partners (the `slumberjacks') are given the following exchange-breaking soft-masses:
\bea
V_\text{Soft} & \simeq & \widetilde{m}^2 \bigg( \big|{\widetilde{Q}}\big|^2 + \big|\widetilde{U}^c\big|^2 - \big|\widetilde{Q}'\big|^2 - \big|\widetilde{U}'^c\big|^2   \bigg) ~~.
\label{eq:VSoft}
\eea
We will show how one can obtain this SUSY breaking pattern from the UV, which will make concrete the interpretation of $\widetilde{m}^2$ as a spurion for the simultaneous spontaneous breaking of SUSY and the exchange symmetry.

We then compute the masses and use them as input to the Coleman-Weinberg potential, from which we obtain the Higgs mass corrections from both the top-stop and lumberjack sectors:
\begin{align}
\delta M_{H_u}^2& =   -\frac{3\, \lambda_t^2}{8\s \pi^2}\, \widetilde{m}^2 \bigg[ R+(R-1)^2 \log(R-1) \nonumber \\
& \hspace{68.5pt} -(R-2)\, R\s \log (R) \bigg] ~~.
\label{eq:quad}
\end{align}
where $R = M^2/\widetilde{m}^2$.  
Due to the presence of the lumberjacks, the UV-sensitivity has been logged away, leaving behind only the IR contribution, as expected.  
To leading order in $1/R$ we have
\be
\delta M_{H_u}^2 \simeq   -\frac{3\, \lambda_t^2}{8\s \pi^2}\, \widetilde{m}^2\s \bigg[ \frac{3}{2}+ \log \left( \frac{M^2}{\widetilde{m}^2} \right) \bigg] ~~,
\ee
which makes clear that, up to an additional finite threshold correction, the scale $M$ effectively replaces the UV-cutoff, which is consistent with the simple renormalization group evolution interpretation of \Fig{fig:idea}. 

More generally, one might imagine extending the third generation matter to include a variety of fields $\boldsymbol{\Phi}_i$ with diverse Standard Model charges and couplings to the Higgs,
\begin{align}
\boldsymbol{W} = \frac{1}{2}\, \lambda_H^{ij}\, \boldsymbol{H_u}\, \boldsymbol{\Phi}_i\, \boldsymbol{\Phi}_j + \frac{1}{2}\, M^{ij}\, \boldsymbol{\Phi}_i\, \boldsymbol{\Phi}_j~~,
\end{align}
as well as soft terms
\begin{align}
V_\text{Soft} =  \frac{1}{2}\, \big(\widetilde m^2\big)_i^j \,\widetilde{\Phi}^{* i } \, \widetilde{\Phi}_j ~~.
\end{align}
The general condition for screening UV-sensitive logs at one loop is simply
$\lambda_{Hki}\, \lambda_H^{kj}\, (\widetilde m^2)^i_j = 0$. When all fields couple with equal strength to the Higgs, this reduces to the simple condition that the trace of the soft masses vanishes, $(\widetilde m^2)^i_i = 0$; this is clearly satisfied by the soft terms in \Eq{eq:VSoft}.

\subsection*{Fine-tuning}
As in the MSSM, to realise the observed Higgs mass one would likely need additional contributions to the Higgs quartic, such as radiative corrections from $A$-terms or by extending the framework to include additional NMSSM-like singlets. 
As a result, any exploration of the parameter space realising the observed Higgs mass is inherently model-dependent.  
However, independent of these considerations, we may still study the \emph{improvement} in fine-tuning relative to a standard MSSM-like scenario.  
Whenever the fine-tuning is dominated by the presence of a heavy stop in the MSSM, the quadratic corrections to the up-type Higgs mass give the leading contribution:
\be
\delta M_{H_u}^2 =   -\frac{3\, \lambda_t^2}{8\s \pi^2}\, \widetilde{m}^2 \log \bigg(\frac{\LUV^2}{\widetilde{m}^2} \bigg)  ~~,
\label{eq:quadMSSM}
\ee
which must be tuned against other contributions, such as the $\mu$-term.  
On the other hand, the corrections in the logged model are given in \Eq{eq:quad}.  
As a result, regardless of how the required quartic is generated, if we simply assume that the fine-tuning is dominated by the stop sector, we can estimate the gain in fine-tuning by computing the ratio of these two corrections as a function of the mass of the lightest coloured scalar.

\begin{figure}[t]
\includegraphics[width=0.5\textwidth]{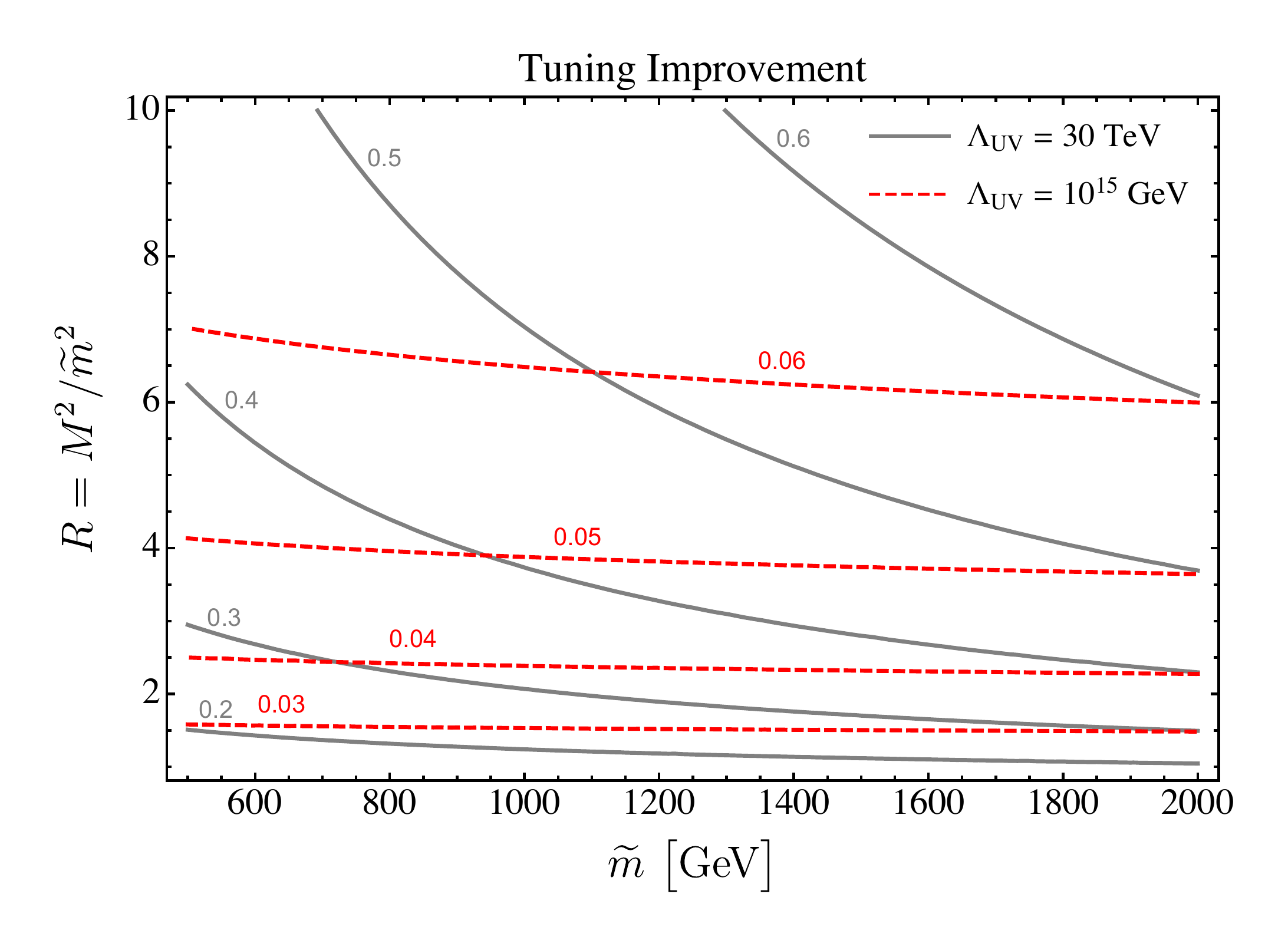}
\caption{Contours of the Higgs mass squared corrections in the logged model (\Eq{eq:quad}) divided by the corrections in the MSSM (\Eq{eq:quadMSSM}), in the $R =  M^2/\widetilde{m}^2$ versus $\widetilde{m}$ plane.  
The gray contours are for $\LUV = 30$ TeV and red for $\LUV = 10^{15}$ GeV, which is relevant for the MSSM corrections.  
Thus these contours represent the \emph{reduction} in fine-tuning in the supersoft model relative to the MSSM.  
We find improvement factors as large as $\sim 30$ for high mediation scales.}
\label{fig:tuning}
\end{figure}

In \Fig{fig:tuning}, we show contours of the improvement in the tuning when the lumberjacks are present as compared to the MSSM in the $R$ versus $\widetilde{m}$ plane.  
Even for low mediation scales, it is clear that the reduction in the tuning is considerable.  
For instance, if one has a GUT-scale model compatible with current bounds on stop squarks with tuning of $\mathcal{O}(1\%)$, then the supersoft version could be realised with a rather negligible tuning of $\mathcal{O}(30\%)$ for low $R$.  Indeed, fine-tuning in logged models approaches the favourable level obtained if one used only the infrared values of supersymmetry-breaking parameters \cite{Baer:2013gva}.

Although this analysis is simplistic, it should capture the leading fine-tuning aspects, though effects beyond leading logarithm are likely to moderate the improvement \cite{Casas:2014eca, Buckley:2016tbs}.  
Looking forward, it would be interesting to revisit the current fine-tuning in supersymmetric models with high scale mediation for a logged model to quantitatively assess the level of pressure current null LHC results put on supersymmetric naturalness.
The gains should be particularly striking if the mediation scale is high.

It bears noting that the radiative correction to the Higgs quartic from the lumberjack fields is negative (since the fermionic lumberjacks are heavier than the scalars) and proportional to $\log M/\widetilde{m}$.
This reduces the overall contribution from the top sector in close analogy with the reduction of the $D$-term quartic in supersoft theories with Dirac gauginos.
A more rigorous treatment of the parameter space would reveal that there is a tradeoff between the reduction in the log contribution to the fine tuning (approximately $\log \Lambda_\text{UV}/M$) and the decreased quartic when the observed Higgs mass is driven by the top sector. 
In this case, the naturalness improvement will be significant as long as $ \widetilde m\, \LUV \gg M^2$.  %We take this as further motivation for the NMSSM-like extension introduced in the next section.

%One possibility is to realise the SM gauge group $G$ as arising from a diagonal component after spontaneously breaking $G_A\times G_B \to G$.  In this way the Higgs couples to a heavy vector sector $\widetilde{G}$ in addition to the SM gauge group.  If gaugino masses are such that SM gauginos are lifted and $\widetilde{G}$ gauginos are lighter than $\widetilde{G}$ gauge fields, then in principle one may realise an analogous structure as for the matter sector.  However, since the tuning in the EW sector is still rather mild we refrain from exploring this direction further.

%\begin{figure*}[t]
%\leavevmode
%\vspace{-.30cm}
%\begin{center}
%\includegraphics[width=0.9\textwidth]{Figs/Scalars.pdf}
%\end{center}
%\vspace{-0.50cm}
%\caption{Currently a place holder (just copied the fermions figure)}
%\label{fig:scalars}
%\end{figure*}

\section{UV-Completions}
\label{sec:complete}
Looking towards the UV, we assume that the SUSY breaking soft masses arise from a superfield $D$-term $\langle \boldsymbol{\Phi} \rangle = D_\Phi\, \theta\strut^2\, \overline{\theta}\strut^2$ which is odd under the exchange symmetry, $\boldsymbol{\Phi} \to -\boldsymbol{\Phi}$.  
The K\"ahler potential at the matching scale $\LUV$ is
\be
\boldsymbol{K}  = \frac{\boldsymbol{\Phi}}{\LUV^2} \Big(\, \big|\boldsymbol{Q}\big|^2 + \big|\boldsymbol{U^c}\big|^2 - \big|\boldsymbol{Q'}\big|^2 - \big|\boldsymbol{U^{\prime c}}\big|^2 \Big) ~~.
\label{eq:coupling}
\ee
Finiteness of the Higgs mass corrections is simply a consequence of the fact that the Higgs cannot couple to the SUSY breaking field in the K\"ahler potential as $K \supset \boldsymbol{\Phi}\, |\boldsymbol{H}|^2 + \text{h.c.}$, since the global $\mathcal{Z}_2$ symmetry forbids any counterterm at this order.

While the EFT is self-consistent it is interesting to consider the possible UV structure.  
Perhaps the simplest possibility would be to realise $\boldsymbol{\Phi}$ as a pair of partner fields which swap under the exchange symmetry
\be
\boldsymbol{\Phi} = \big|\boldsymbol{X}\big|^2 - \big|\boldsymbol{\overline{X}}\big|^2 ~~.
\label{eq:leading}
\ee
The sign enforces consistency with the coupling in \Eq{eq:coupling}.
An $F$-term of this form could be generated in a number of ways.  
The simplest possibility would be to have $\boldsymbol{W_X} = f\s \boldsymbol{X}$, although completions of this scenario may require more involved superpotentials to render all additional fields massive.
Another viable approach would be to utilize an O'Raifeartaigh-like superpotential \cite{ORaifeartaigh:1975nky} with a small source of exchange symmetry breaking to yield the desired $F$-term.

It is also important to understand the impact of additional operators that could be present in the UV.  
Perhaps the most concerning coupling is 
\be
\boldsymbol{K_H} = \frac{\big|\boldsymbol{X}\big|^2 + \big|\boldsymbol{\overline{X}}\big|^2}{\Lambda_H^2}\, \big|\boldsymbol{H}\big|^2 ~~,
\label{eq:KHBad}
\ee
which is allowed by the symmetries and would spoil the supersoftness.  
Therefore, our IR analysis is valid under the assumption that the leading coupling is to the matter fields, as in \Eq{eq:coupling}.
Then the generation of the dangerous coupling in \Eq{eq:KHBad} would be suppressed by extra powers of $\widetilde{m}^2/\LUV^2$, or by some number of loop factors.
Both of these contributions probe the dynamics of the messenger and SUSY-breaking sectors, and are consequently highly UV dependent.

An alternative UV scenario could be constructed by introducing an additional anomalous $\U(1)_X$ gauge symmetry under which the MSSM and lumberjack superfields carry equal and opposite charges.  
These charge assignments are consistent with the exchange symmetry if the vector superfield is odd $\boldsymbol{V} \to -\boldsymbol{V}$.  
The $\U(1)_X$ can be spontaneously broken supersymmetrically if a pair of superfields $\boldsymbol{X}$ and $\boldsymbol{\overline{X}}$ with equal and opposite charge have vevs.  
Integrating out the vector multiplet could be responsible for generating \Eq{eq:coupling}.
Since these fields also transform under the exchange symmetry, they may be used to generate the exchange-symmetry-respecting superpotential Yukawa couplings given in \Eq{eq:Wlambda}.  
The exchange-breaking superpotential masses in \Eq{eq:WM} may be generated by a small exchange-breaking Yukawa coupling involving $\boldsymbol{X}$ and the lumberjacks.  
Finally, the construction of \cite{ArkaniHamed:1998nu}, wherein an anomalous $\U(1)_X$ symmetry obtains a supersymmetry-breaking $D$-term well below the scale of spontaneous gauge symmetry breaking, could provide the source of the soft masses in \Eq{eq:VSoft}.
These two simplistic scenarios suggest that a comprehensive UV-completion is feasible, although it is likely that it will need to rely on some non-generic features.

\subsection*{Gauginos}
In the lumberjacking spirit, we may also opt to have Dirac gauginos, since this generates one-loop finite radiative corrections in the scalar sector \cite{Fox:2002bu}.  
This can be realised consistently with the existing setup since the adjoint chiral superfields $\boldsymbol{A}_j$ may be made odd under the exchange symmetry, such that a Dirac gaugino mass may be generated through the usual operator
\be
\boldsymbol{W}_{\text{Gauge}} = \frac{1}{\LUV} \boldsymbol{W^\Phi}_\alpha\s \boldsymbol{W}^\alpha_j\s \boldsymbol{A}_j ~~,
\ee
where $\boldsymbol{W}^\Phi_\alpha$ is the vector chiral superfield for $\boldsymbol{\Phi}$ whose $D$-term vev yields the Dirac mass, and $j$ denotes the choice of gauge group.
The $\mathcal{Z}_2$ symmetry acts on the adjoint chiral superfield as $\boldsymbol{A}_j \to -\boldsymbol{A}_j$.  
Since the IR exchange symmetry is respected by the gauge sector, the arguments for the scalar sector given in the previous section are unmodified.
The dangerous supersoft-spoiling mass terms associated with the $\mu-B_\mu$ problem of Dirac gaugino scenarios may be avoided through the GoGa mechanism \cite{Alves:2015kia,Alves:2015bba}.  

\subsection*{Higgs Quartic}
To raise the Higgs quartic without introducing UV-sensitive logarithms we may also supersoften the singlet sector of the NMSSM.  To do this we follow the spirit of the top sector and introduce two singlets with an exchange symmetry
\begin{align}
\boldsymbol{W_S} = \lambda\, \left( \boldsymbol{S} + \boldsymbol{S'} \right)\, \boldsymbol{H_u}\, \boldsymbol{H_d} + \frac{1}{2} M_{\boldsymbol{S}} \left( \boldsymbol{S}^2+ \boldsymbol{S'}^2  \right) ~~.
\end{align}
If desired, additional cubic interactions can be added consistent with the exchange symmetry.  To introduce SUSY breaking we include an exchange-breaking soft term
\bea
V_\text{Soft} & \simeq & -\widetilde{m}_S^2 \bigg(  |\widetilde{S}|^2 -  |\widetilde{S}'|^2 \bigg) ~~,
\label{eq:VSoft}
\eea
with a similar origin to the soft masses in the stop sector.  As before, the one-loop radiative corrections will be IR-calculable, allowing to raise the Higgs mass without the concern of large logarithms.  Furthermore, since in the supersoft model the quadratic corrections tend to be smaller than standard scenarios with even a very low messenger scale, it is likely that the fine-tuning outcome for the singlet sector will be less severe than in standard NMSSM scanerios \cite{Barbieri:2006bg}.

\section{Phenomenology}
\label{sec:pheno}
Much of the LHC phenomenology is driven by the Dirac nature of the gauginos.  Studies of the novel signatures of Dirac gauginos have been undertaken previously \cite{Choi:2008ub, Choi:2009ue, Blum:2016szr, Alvarado:2018rfl}.  
Due to the supersoftness, the gluinos can be heavier than in the MSSM without being the dominant source of tuning,   which implies that jets plus missing energy signatures from the gluino and light squark sector have a lower rate, see~\emph{e.g.}~\cite{Kribs:2012gx}.  

The novel feature of this model is the presence of the vector-like lumberjacks.  Note that in the model as described in \Sec{sec:Scalars}, these fields would be stable and hence, depending on the cosmological history, potentially phenomenologically unacceptable.  To complete the story, one could introduce small $\mathcal{Z}_2$ breaking terms that enable lumberjack decays.  Since this coupling may violate the symmetry responsible for removing the logs, these couplings may reintroduce logarithmic UV-sensitivity.  Therefore, it is necessary that they small enough to not spoil the reduction in tuning.  While it is not required, it is interesting to consider the region of parameter space where these couplings are so small that the decays are displaced.  This could be motivated by models where the $\mathcal{Z}_2$ breaking is generated at the UV scale by a higher dimension operator. 

The collider phenomenology of a supersymmetric vector-like fourth generation has been studied previously (see, for example, \cite{Martin:2009bg,Graham:2009gy}), and is rich.  In particular, logging requires the particular slumberjack soft-mass pattern given in \Eq{eq:VSoft}.  Post discovery, detailed measurements of the mass spectrum could yield strong evidence that this mechanism is being used by nature to avoid excessive fine-tuning of the weak scale.  It would be interesting to study the viability of such measurements at the HL-LHC or FCC-hh.  However, since the detailed signatures are highly model dependent in a manner that is unconnected to the supersoft properties we leave the exploration of this phenomenology to future work.

\section{Summary}
\label{sec:sum}
Naturalness seeds of doubt were sown when LEP measurements demonstrated precision agreement with the predictions of the Standard Model.  They have since germinated due to the paucity of discoveries beyond the Higgs boson at the LHC and are now maturing into a naturalness crisis \cite{Giudice:2017pzm}.  Our hope to experimentally access the microscopic provenance of the electroweak scale has been challenged by the interpretation of null results in the context of classic scenarios such as the MSSM and minimal composite Higgs models.

When one assumes that  the electroweak hierarchy problem is tamed through the introduction of a new symmetry that commutes with the Standard Model gauge groups, there is a generic expectation that new coloured `top-partner' states should exist in proximity to the weak scale.   Since LHC null-results currently imply $\widetilde{m} \gtrsim 900$ GeV, the resultant na\"ive scaling of Higgs mass corrections (\Eq{eq:quadMSSM})
%\be
%\delta M_H^2 \propto \frac{\widetilde{m}^2}{16\s\pi^2} \log \bigg(\frac{\Lambda}{\widetilde{m}}\bigg)
%\label{eq:scaling}
%\ee 
might cause one to conjecture that naturalness is not a useful guide for predicting the next layer of fundamental physics.  This is exacerbated if one assumes that the fundamental microscopic scales satisfy $\LUV \gg M_H$ as this raises the Higgs mass parameter through large logarithmic enhancements.  However, we emphasise here that this scaling is only a minimal expectation, subject to very basic underlying assumptions.  

To this end we have embraced the log-enhanced contributions and, rather than attempting to banish them by reducing $\LUV$, we have instead introduced the lumberjack fields, who screen logs through equal and opposite-sign contributions from the mediation scale $\LUV$ down to the scale $M$ where they decouple.  This significantly reduces the fine-tuning associated with the top sector --- resulting in a model where a mediation scale $\LUV \gg M_H$ can be just as natural as one where the mediation scale is low.  

This supersoft strategy has been previously studied for the gauge sector, and is shown here to be possible for the matter sector.  In the models presented here, one may have a GUT-scale mediation scale and stops heavier than $\sim 1$ TeV, consistent with current limits \cite{Aaboud:2017aeu, Sirunyan:2019glc}, with fine-tuning reduced by a factor $\mathcal{O}(30)$ as compared to the most basic MSSM expectation.
%Care should be taken not to over-interpret the quantitative value of any fine-tuning measure in a specific model, however fine-tuning measures do afford an IR inhabitant a qualitative feeling for how `likely' different UV scenarios may be, especially when all corrections are fully calculable, as in this supersoft stop model.  

The reduced fine-tuning in this supersoft stop model makes clear it is dangerous to over-interpret naturalness implications of null results from LHC coloured particle searches, since any interpretation is highly model dependent.  Nature does not have to make the minimal choice, thus it is premature to conclusively declare that high-scale SUSY is incompatible with naturalness.

%%%%%%%%%%%%%%%%%%%%%%%%%%%%%%%%%%%%%%%%%%%%%%%%%%%%%%%%%%%%%%%%%%%%%%
\acknowledgments{
This work is dedicated to the original supersoft pioneer, Ann Nelson.  TC, NC, SK, and MM are grateful to the KITP `Origin of the Vacuum Energy and Electroweak Scales' program, where this work was initiated. This research was supported in part by the National Science Foundation under Grant No. NSF PHY-1748958.  TC is supported by the U.S. Department of Energy (DOE), under grant DE-SC0011640.  MM is supported by the STFC HEP consolidated grant ST/P000681/1 and Trinity College Cambridge. NC and SK are supported by the U.S. Department of Energy (DOE) under the Early Career Award DE-SC0014129 and the Cottrell Scholar Program through the Research Corporation for Science Advancement.  JTS is supported by the STFC consolidated grant ST/S505316/1.
}

%%%%%%%%%%%%%%%%%%%%%%%%%%%%%%%%%%%%%%%%%%%%%%%%%%%%%%%%%%%%%%%%%%%%%%

%\end{spacing}
%%%%%%%%%%%%%%%%%%%%%%%%%%%%%%%%%%%%%%%%%%%%%%%%%%%%%%%%%%%%%%%%%%%%%%
%%%%%%%%%%%%%%%%%%%%%%%%%%%%%%%%%%%%%%%%%%%%%%%%%%%%%%%%%%%%%%%%%%%%%%
%\appendix
%%%%%%%%%%%%%%%%%%%%%%%%%%%%%%%%%%%%%%%%%%%%%%%%%%%%%%%%%%%%%%%%%%%%%%
%%%%%%%%%%%%%%%%%%%%%%%%%%%%%%%%%%%%%%%%%%%%%%%%%%%%%%%%%%%%%%%%%%%%%%

%%%%%%%%%%%%%%%%%%%%%%%%%%%%%%%%%%%%%%%%%%%%%%%%%%%%%%%%%%%%%%%%%%%%%%
%%%%%%%%%%%%%%%%%%%%%%%%%%%%%%%%%%%%%%%%%%%%%%%%%%%%%%%%%%%%%%%%%%%%%%
%%%%%%%%%%%%%%%%%%%%%%%%%%%%%%%%%%%%%%%%%%%%%%%%%%%%%%%%%%%%%%%%%%%%%%
\bibliographystyle{JHEP}
\bibliography{references}
%%%%%%%%%%%%%%%%%%%%%%%%%%%%%%%%%%%%%%%%%%%%%%%%%%%%%%%%%%%%%%%%%%%%%%
%%%%%%%%%%%%%%%%%%%%%%%%%%%%%%%%%%%%%%%%%%%%%%%%%%%%%%%%%%%%%%%%%%%%%%

\end{document}